\documentclass[letterpaper,12pt]{article}

\usepackage[utf8]{inputenc}
\usepackage{url}
\usepackage{microtype}
\usepackage{graphicx}
\usepackage{booktabs}
\usepackage{tabulary}
\usepackage{subcaption}
\usepackage[margin=2cm]{geometry}


\usepackage[unicode=true]{hyperref}
\hypersetup{breaklinks=true,
            bookmarks=true,
            pdfauthor={E. Graells-Garrido and B. Poblete},
            pdftitle={\#Santiago is not \#Chile, or is it? A Model to Normalize Social Media Impact},
            colorlinks=true,
            urlcolor=blue,
            linkcolor=magenta,
            pdfborder={0 0 0}}

\title{\#Santiago is not \#Chile, or is it? A Model to Normalize Social Media Impact}
\author{
Eduardo Graells-Garrido \\
Web Research Group \\
Universitat Pompeu Fabra \\
Barcelona, Spain \\
\texttt{eduard.graells@upf.edu}
\and
B\'arbara Poblete \\
PRISMA Research Group \\
Depto. de Ciencias de la Computaci\'on \\
Universidad de Chile \\
Santiago, Chile \\
\texttt{bpoblete@dcc.uchile.cl} 
}

\date{}

\begin{document}
\maketitle

\begin{abstract}
Online social networks are known to be \emph{demographically biased}. Currently there are questions about what degree of representativity of the \emph{physical population} they have, and how population biases impact user-generated content. In this paper we focus on \emph{centralism}, a  problem affecting Chile.
Assuming that local differences exist in a country, in terms of vocabulary, we built a methodology based on the vector space model to find distinctive content from different locations, and use it to create classifiers to predict whether the content of a micro-post is related to a particular location, having in mind a geographically diverse selection of micro-posts.
We evaluate them in a case study where we analyze the virtual population of Chile that participated in the Twitter social network during an event of national relevance: the municipal (local governments) elections held in 2012.
We observe that the participating \emph{virtual population} is spatially representative of the \emph{physical population}, implying that there is  centralism in Twitter. Our classifiers out-perform a non geographically-diverse baseline at the regional level, and have the same accuracy at a provincial level. 
However, our approach makes assumptions that need to be tested in multi-thematic and more general datasets. We leave this for future work.

\end{abstract}

%\category{H.3.3}{Information Storage and Retrieval}{Information Search and Retrieval}[Information Filtering]

%\terms{Social Networks, Population, Geography}

%\keywords{Text geolocation, Vector Space Model, Population Characterization}

\section{Introduction}
\label{sec:intro}

Chile is known as a country with a high participation of its population in social networks on Internet: it has the highest penetration of Facebook\footnote{\url{https://facebook.com}, visited 17-July-2013/.} in Latin America \cite{socnetlatam}. 
Undoubtedly, this active presence in social networks has changed the way people communicate and how media and government relate with people. For instance, media outlets use social media to interact with people and find out what their interests and opinions on a certain topic are.
Although Facebook is the social network with most penetration in Chile, another social network, Twitter \footnote{\url{https://twitter.com}, visited 10-July-2013.}, has much more presence in the public sphere, given its public nature. Conversations and micro-posts, also called \emph{tweets}, are inherently public (although it is possible to make them private, only a small fraction of accounts use this option). Therefore, to browse the content published in Twitter an account is not needed, and a search by keywords or topics, annotated through hashtags, is all anyone needs to be able to stay up-to-date with respect to what is trending at that moment. 

However, in the \emph{physical world}, Chile is becoming more centralized everyday \cite{wiki:chilecentralism}, and it is not clear if the virtual population present in Twitter reflects or is representative of the physical population, and to what degree this virtual population is also centralized.
A common saying is ``Santiago is not Chile''\footnote{Santiago is the capital of Chile.}, referring to the fact that the capital is not representative of the country, yet media outlets concentrate in Santiago and government policies are tailored at the needs of Santiago. Given the climatic, geographical and cultural diversity of Chile, centralism is a serious problem. One example is that the law that dictates minimum housing requirements is the same for all regions of the country \cite{leyvivienda}, in spite of the extreme weather differences between northern and southern regions.

Previous literature regarding this subject states that more populated cities are over-represented in Twitter, while less populated cities are under-represented \cite{mislove2011understanding}. Because of over-representation of populated places, in addition to the lack of balance in physical population distribution in Chile, content from or about non largely populated locations is lost in the timeline of tweets. In addition, it is hard to find local content, as the only salient ways to find content are to click on trending topics, which by definition are biased towards more populated cities, and by searching. But this implies that the information seeker already knows what to look for, if an information need effectively exists. As such, there is no current way to explore a geographically diverse timeline: users have the responsibility to follow diverse accounts. However, current interfaces do not allow users to see a diversity of tweets according to any criteria. Since Twitter recommends users (the \emph{``who to follow''} functionality) and tweets (the \emph{``discover''} tab) based on account connections and activity. Users who do not have diverse connections will not receive diverse recommendations.

Motivated by the situation described, in this paper we propose a methodology to address the following research questions:

\begin{itemize}
  \item \textbf{Q1:} Does the participating virtual population (in an event in Twitter) represent the physical population, and to what degree?
  \item \textbf{Q2:} Given the content generated by the virtual population, is it possible to identify local content using a lexical approach?
\end{itemize}

Our work presents the following contributions: \emph{1)} based on the vector space model, we build language models and classifiers that, given a small text, such as a tweet, predict the location the text is talking about. This is a different approach from previous work that focused on classifying the location of users; \emph{2)} a case study of the virtual population of Chile in Twitter. In particular, we address the virtual population that participated in a national event of high local importance: the municipal elections which took place in Chile in October 28th, 2012 \cite{wiki:chile2012elections}. We apply our methodology in this dataset and evaluate our classifiers.

We discuss and conclude that, having characterized the participating virtual population in the event, its distribution reflects the distribution of the physical population of Chile, and we find that it is possible to identify content related to different locations using our methodology. In addition, our classifiers allow us to build geographically diverse timelines with equal or better accuracy than a non diverse baseline.

\section{Related Work}
There is no clear answer to the question \emph{what is Twitter?} \cite{kwak2010twitter}. However, a wide spectrum of research areas have seen it from different perspectives. 
One of them is the geographical span of networks: previous work has found that, the stronger the network (defined in terms of reciprocity in connections, 1-way and 2-way interactions by mentioning others), the lower is its geographical span \cite{quercia2012social}. In terms of discussion, local events have more dense networks of discussion than global events, and central individuals in the network are also located centrally in the physical world \cite{yardi2010tweeting}. 
A demographic study of user accounts from the the U.S.A. concluded that populated cities are over-represented, while less populated cities are under-represented in Twitter \cite{mislove2011understanding}. 

To understand population in Twitter is necessary to determine the location of a person, a meta-attribute not directly available from user profiles. Different methods have been proposed to \emph{geolocate} users. From simple detection based on the free-text self-reported location \cite{mislove2011understanding} to probabilistic language models to find out if the language used by a person denotes its location \cite{hecht2011tweets,kinsella2011m,o2013modeling}. Although complex models achieve higher accuracy than simpler models, they also require a representative and often massive corpus of users and locations. Our population characterization is based on the self-reported location, which contains geographical information in $66\%$ of the cases \cite{hecht2011tweets}. 

Interactions at the global level, in the sense of interplay between countries, is also an area related to our work. Countries tend to communicate and form clusters based on language, and geography plays a big role in interacting with others at the country level \cite{kulshrestha2012geographic}. 
It is possible to study how different countries, which have different cultures, have different behaviors in terms of tweeting frequencies, interactions and network structure \cite{poblete2011all}. In addition, tweeting behavior correlates to actual culture metrics: \emph{pace of life}, \emph{power distance} and \emph{individualism} \cite{ICWSM136102}. In our work, we assume that local differences exist, and we take this into account to define our methodology and perform our analysis.

\section{Methodology}

Twitter is a social network where users post status updates with a maximum length of 140 characters, called \emph{tweets}. Connections between users are directed, and when user A \emph{follows} user B, tweets and \emph{re-tweets} made by B will show-up in A's timeline. The timeline is a list of tweets in reverse chronological order, as it is expected that a twitter user is interested in what is happening at the present time. To find previously posted content, users can search or go back in the timeline, but in both cases there are time and content limits on what can be found. Users can annotate tweets using \emph{hashtags}, keywords that start with the hash character \texttt{\#}. Twitter supports its usage by auto-linking a hashtag to its correspondent search results.

\textbf{User Locations}: To geolocate users, we relied on the self-reported location in users' profiles. Instead of querying external services for geolocation using profile locations as input, as in \cite{mislove2011understanding}, we chose to build a list of common location names. In \cite{hecht2011tweets} location names extracted from Wikipedia are used. Our approach is similar, but instead of relying on the list provided by Wikipedia, we use \emph{templates} to generate a list of valid location names.

\textbf{Vector Space Model}: 
We represent groups of tweets as documents in the \emph{vector space model} \cite{salton1975vector,baeza2011modern}, where the position $i$ in the document vector represents the weight of the word $i$ from the vocabulary according to its frequency in the document (\emph{TF}, \emph{term frequency}), normalized by its \emph{inverse document frequency} (\emph{IDF}), the number of documents in which the word appears. This schema, known as TF-IDF, allows to discard corpus specific stopwords by assigning lower weights to words that appear in many documents, and allows to discover discriminating words that appear in fewer documents.

\subsection{Geolocating Tweets}
\label{sec:classifiers}

It is possible that users from different locations are aware of how timelines are biased towards more populated locations, and thus use local hashtags that can help them and other interested persons to find local content. We assume the presence of uniquely local hashtags that indicate the location a tweet is reffering to.
Therefore, there should be a way to predict if a tweet talks about a location, even on the absence of local relevant hashtags obtained with TF-IDF: words that co-occur with locally relevant hashtags may also be of local usage. Its association with the identified local hashtags and keywords can be used to geolocate users using a classifier built on language models. In previous work, this approach has yielded good results when the input dataset is: \emph{1)} big enough to faithfully represent the different locations to predict, and \emph{2)} diverse, covering different topics and events. 
Our approach aims at a different scenario, where the dataset is small to medium sized. Moreover, our focus is on singular events, and thus complex approaches might not have enough data to build representative language models. This has motivated us to use a simpler approach: our tweet classifiers use TF-IDF weighting for documents (groups of tweets). The grouping criteria depends on the approach (see Section \ref{sec:case_study_classifiers}).

Our classifiers work by projecting documents into the vector space defined. To project a document into a TF-IDF space, the \emph{bag of words} vector of a document is weighted according to term frequencies and inverse document frequencies of its words in relation to the documents used to build the model.

Additionally, to test more complex approaches in our scenario, we use \emph{Latent Semantic Indexing} \cite{deerwester1990indexing}. Again, we use groups of tweets as documents. We build a term-document $t \times d$ matrix $M$, where $M_{i,j}$ denotes the TF-IDF weight of term $i$ in document $j$. Then, we do a \emph{singular value decomposition} (SVD) to decompose the matrix into:
$$
M = T \times S \times D^{T}
$$
where $T$ is a matrix of $t \times r$ with orthogonal columns, $S$ is a $r \times r$ diagonal matrix of decreasing singular values, and $D$ is a $d \times r$ matrix of document vectors. Then, the $S$ matrix is truncated to $k < r$, where $k$ is the number of latent dimensions or latent topics. Optimal values of $k$ are usually between the range $200$--$2000$. The projection of a vector $v$ is defined as $v \times T \times S^{-1}$. 

After building the TF-IDF and LSI language models, we project documents for each location with all their corresponding tweets into each model, and then save the projected documents for later comparison with a query. Given a query, we also project its text into the model space, and then we compute the cosine similarity between the projected query $q$ and each projected location document, $L$:
$$
cosine\_similarity(q, L) = \frac{q \cdot L}{\parallel q \parallel  \parallel L \parallel }
$$
The location with the highest similarity to the query will be predicted as the location of the tweet content.

\section{Case Study}

Our dataset is composed by tweets crawled on October 28th, 2012, related to the municipal elections performed in Chile. The event had a distinctive hashtag, \texttt{\#municipales2012}, which among with other related hashtags, keywords, location and candidate names, were used as queries for the \emph{Twitter Streaming API}\footnote{\url{https://dev.twitter.com/docs/streaming-apis}, accessed 11-July-2013.}. We started to crawl tweets at 10:00 AM and stopped at midnight. Table \ref{table:information_space} shows the overview of the dataset after removing unrelated tweets by manual inspection of non-related keywords and hashtags.

\begin{table}
\centering
\begin{tabular}{lc}
  \toprule
  Data & \# \\
  \midrule
  Tweets & $498594$ ($7.05\%$ with geo. info.)\\
  ReTweets & $270028$ \\
  Accounts & $173077$ \\
  \bottomrule
\end{tabular}
\caption{Our information space: main types of data crawled during the \texttt{\#municipales2012} event.}
\label{table:information_space}
\end{table}

\subsection{Virtual Population}

The administrative locations of Chile are defined according to the following hierarchy: $municipality \rightarrow province \rightarrow region \rightarrow country$. After looking some common location names in the dataset, we defined a list of templates to generate location names: 
1) \emph{municipality},
2) \emph{province},
3) \emph{municipality}, \emph{province},
4) \emph{province}, \emph{region},
5) \emph{region},
6) \emph{municipality}, \emph{country},
7) \emph{province}, \emph{country},
8) \emph{region}, \emph{country},
9) \emph{municipality} de\footnote{\emph{de} is spanish translation of \emph{of}.} \emph{country},
10) \emph{province} de \emph{country},
11) \emph{region} de \emph{country},
12) \emph{country}.

We generated location names from these templates based on the official location names from Chile \cite{minint2010}\footnote{This data is published as a SQL database in \url{https://github.com/knxroot/BDCUT_CL}, accessed on 11-July-2013.}. After detecting user-reported locations compliant to our templates, we inspected the database to search for additional location names that could be assigned to a country, province or region of Chile, or even to Chile itself, and that were not template compliant. At the end of this process we had $1978$ valid location names. Since we covered all locations with more than one account, we believe the usage of external services for geolocation was not needed. 

\begin{table}
\centering
\begin{tabular}{lcc}
\toprule
Level & \# Users & \# Tweets + RTs\\
\midrule
Country & $17929$ ($10.4\%)$        & $98458$ ($12.8\%)$\\
Region & $341$ ($0.1\%)$     & $2637$  ($0.3\%)$\\
Province & $1842$ ($1.1\%)$      & $12541$ ($1.6\%)$\\
City/Municipality & $52961$ ($30.6\%)$ & $325326$ ($42.3\%)$\\
\midrule
Empty & $60581$ ($35.0\%$) & $188533$ ($24.5\%$) \\
N/A & $39423$ ($22.8\%)$ & $141127$ ($18.4\%)$\\
\bottomrule
\end{tabular}
\caption{Number of accounts geolocated using the self-reported location. In N/A we include accounts with valid information location but outside of Chile. Percents were rounded to one decimal.}
\label{table:location_levels}
\end{table}

\begin{figure}[tb]
\centering
\includegraphics[width=0.4\textwidth]{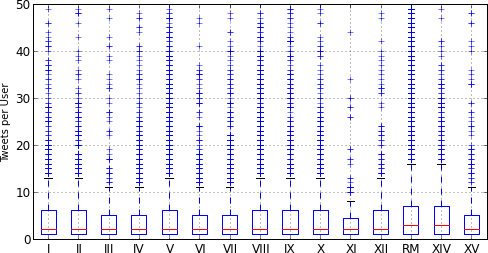}
\caption{Boxplot of population of each region in Chile with number of tweets per geolocated account. Cropped at $30$ tweets per user to showcase the differences and similarities between regions.}
\label{fig:population_region_tweets}
\end{figure}

\begin{figure*}[htb]
\centering
\begin{subfigure}[b]{0.45\textwidth}
        \centering
        \includegraphics[width=\textwidth]{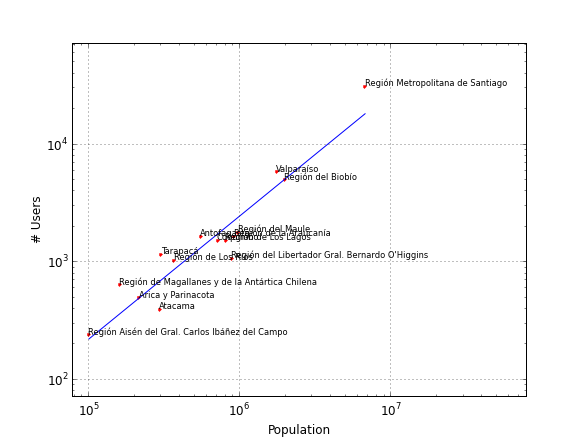}
        \caption{Population of Regions of Chile.}
        \label{fig:population_regions}
\end{subfigure}%
\begin{subfigure}[b]{0.45\textwidth}
        \centering
        \includegraphics[width=\textwidth]{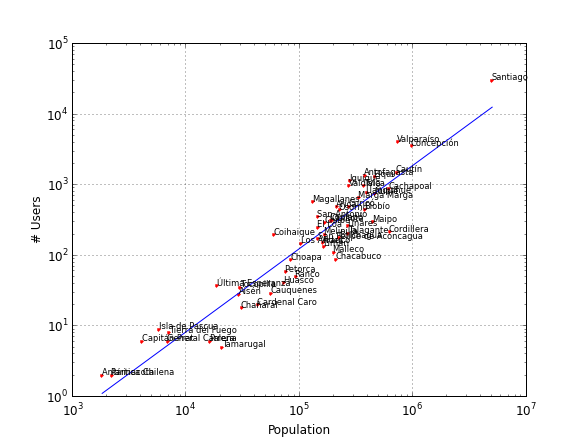}
        \caption{Population of Provinces of Chile.}
        \label{fig:population_provinces}
\end{subfigure}

\caption{Population of regions (left) and provinces (right) and number of accounts per location. The correlation of the logarithms of population and accounts is $0.95$ (regions) and $0.94$ (provinces).}
\label{fig:populations}
\end{figure*}

Table \ref{table:location_levels} contains the number of users and their published tweets per location level. We observe that only a $42.2\%$ of the participating accounts in the event can be localized in Chile. The remaining accounts have empty locations, more than one location, foreign locations or long-tail valid locations not found using our approach. However, the geolocated accounts produce $57.1\%$ of the event content, and as such we consider it a good sample. The logarithms of population and twitter accounts have a pearson correlation of $0.94$ for provinces ($p < 0.01$) and $0.95$ for regions ($p < 0.01 $). 
Figure \ref{fig:populations} shows the population and number of accounts per province (Figure \ref{fig:population_provinces}) and region (Figure \ref{fig:population_regions}). Even though the location populations differ on orders of magnitude (the figures are in log-log scale), the rate of regional twitter accounts per $1000$ inhabitants is $2.61 \pm 0.97$, indicating that in relative terms, the proportion of twitter accounts in each region is similar. The average number of tweets per account is $6.17 \pm 11.53$. The high standard deviation indicates that there are too many accounts with: \emph{1)} only one tweet (the median is $2$),  \emph{2)} a high number of tweets. 
To explore this, in Figure \ref{fig:population_region_tweets} we show the distributions of tweets per account in regions in a boxplot, showcasing that in general the accounts have a similar behavior at each region. Still, there are many outliers that deviate from the upper quartile of tweets per account. For instance, the most active account is a news outlet (\texttt{@cooperativa}) with $479$ tweets.

\subsection{Event Content}

\begin{figure*}[tb]
\centering
\includegraphics[width=0.8\textwidth]{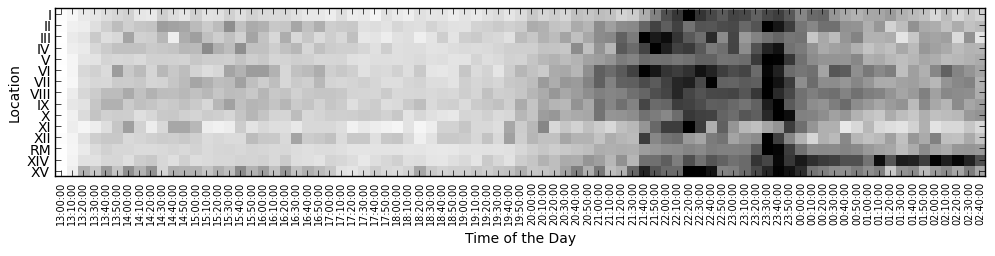}
\caption{Time-series of tweet volume for all regions in Chile, sampled every $10$ minutes. Every region is a row. The darker the shade of each cell, the greater the amount of tweets.}
\label{fig:region_timeseries}
\end{figure*}

Figure \ref{fig:region_timeseries} shows the regional tweet volume during the event. Although the absolute volume of tweets varies greatly between regions, the time-series are fairly similar, showcasing that the event had a common structure at national level.

Table \ref{table:top_hashtags} shows the top $10$ hashtags at national level (of a total of $20509$ hashtags found on the dataset). We observe that half of the top $10$ is related to locations, indicating that location names might be used as cues to indicate to what location a tweet is referring to.

\begin{table}[htbp]
\centering
\footnotesize
\begin{tabular}{l|c|c}
\toprule
Hashtag & \#Tweets & Comment\\
\midrule
\texttt{\#municipales2012} & $115710$ & Event identification.\\
\texttt{\#tudecides} & $19230$ & Event (general)\\
\texttt{\#labbe} & $9565$ & Candidate's name \\
\texttt{\#chile} & $5306$ & Location (country) \\
\texttt{\#nunoa} & $5250$ & Location (municipality) \\
\texttt{\#valdiviacl} & $4701$ & Location (city) \\
\texttt{\#iquique} & $4603$ & Location (city) \\
\texttt{\#providencia} & $4560$ & Location (municipality)\\
\texttt{\#cooperativa} & $3451$ & Media \\
\texttt{\#yovote} & $3015$ & Event (general) \\
\bottomrule
\end{tabular}
\caption{Top $10$ global hashtags at the \texttt{\#municipales2012} event.}
\label{table:top_hashtags}
\end{table}

Table \ref{table:region_relevant_hashtags} shows the $3$ most discriminative hashtags per region obtained with TF-IDF, after discarding hashtags that appear in less than $5$ different tweets. We observe that most of these hashtags are location names inside that region (like \texttt{\#laserena}), names of candidates from the corresponding region (like \texttt{\#karenrojo}), and local adaptations of general hashtags (like \texttt{\#municipalesmag}).

\begin{table}[tb]
\centering
\footnotesize
\begin{tabulary}{0.45\textwidth}{L|L}
\toprule
Region & Relevant Hashtags  \\ \midrule
I & \texttt{\#soria \#altohospicio \#tarapaca} \\ \midrule
II & \texttt{\#karenrojo \#antofagasta \#calama} \\ \midrule
III & \texttt{\#atacama \#copiapo \#freirina} \\ \midrule
IV & \texttt{\#laserena \#coquimbo \#combarbala} \\ \midrule
V & \texttt{\#concon \#quillota \#yovotoucvradio} \\ \midrule
VI & \texttt{\#rancagua \#graneros \#machali} \\ \midrule
VII & \texttt{\#talca \#talcavota \#curico} \\ \midrule
VIII & \texttt{\#chillan \#yolucho \#coronel} \\ \midrule
IX & \texttt{\#araucania \#temuco \#araucaniaelige} \\ \midrule
X & \texttt{\#puertomontt \#osorno \#puertovaras} \\ \midrule
XI & \texttt{\#coyhaique \#aysen \#angol} \\ \midrule
XII & \texttt{\#puqvota \#puq \#municipalesmag} \\ \midrule
RM & \texttt{\#macul \#melipilla \#chilevotaust} \\ \midrule
XIV & \texttt{\#valdiviacl \#uach \#valdivia} \\ \midrule
XV & \texttt{\#aricavota \#arica \#votaarica} \\ \bottomrule
\end{tabulary}
%\end{tabular}
\caption{Top $3$ discriminating hashtags per region, using TF-IDF weighting for tweets grouped by region.}
\label{table:region_relevant_hashtags}
\end{table}

\subsection{Classifiers}
\label{sec:case_study_classifiers}

Only a $31.57\%$ of tweets in our dataset have hashtags. Thus, searching for local hashtags would allow users to find less than one third of the tweets related to the event. 
We built our classifiers by creating documents in the vector space model, using TF-IDF weighting. The definition of document depends on the approach used. Currently we define four approaches:

\begin{itemize}
  \item \textbf{TF-IDF U}: gather all tweets by geolocated users, and create one document per user with their corresponding tweets. 
  \item \textbf{TF-IDF L}: gather all tweets posted from locations, and create one document per location with their corresponding tweets.
  \item \textbf{TF-IDF H}: pick the top $1\%$ used hashtags (discarding the top one because it is too general), and for each hashtag create a document with all tweets containing it.
  \item \textbf{LSI U}: as in TF-IDF U, we use user-documents to build the term-document matrix. Since we have few documents, we use $k = 200$ as the number of latent dimensions.
\end{itemize}

\subsection{Evaluation}

To evaluate the classifiers we use a $10$-fold stratified cross-validation. We divided the set of tweets from geolocated users in $10$ groups, maintaining the proportions of location tweets in each group, and then run $10$ iterations to evaluate the classifiers. In each iteration we built the models using $9$ groups and test the predictions with the remaining group. In this way, each document is used $9$ times for training and one time for evaluation. We do not consider retweets and replies to avoid duplicate tweets in training and evaluation sets.

Since we consider only geolocated users, we assume that most tweets from an account are about its location. We define that a prediction of tweet location is correct if the location predicted for the tweet content (without considering tweet or author meta-data) matches the author location. This approach may give \emph{false positives} (i.e., when a user tweets about other locations) or \emph{false negatives} (i.e., when a user tweets about the event from a generic point of view), but in both cases it is the best approximation we know of. This assumption is also made in previous work.

We test our classifiers at two levels: provinces and regions. We consider all $15$ regions, but only $27$ (of $54$) provinces -- those with equal or more tweets than the median of tweets per province ($936$), as some provinces do not have enough tweets to perform the $10$-fold separation or to build a reliable model. 
Table \ref{table:evaluation} shows the average accuracy of our classifiers, defined as $n_{c}/N$, where $n_{c}$ is the number of correct predictions and $N$ is the total number of tweets. The baseline is a trivial classifier that assigns the most common location in terms of tweets to all predictions: \emph{Santiago} at province level, and \emph{Regi\'on Metropolitana} at regional level. 
At provincial level, \texttt{TF-IDF L} has the same performance as the baseline.
At regional level, \texttt{TF-IDF L} and \texttt{TF-IDF H} outperform the baseline. It is expected that a wider geographical span has better results as each location document is more diverse and complete.   
The \texttt{TF-IDF L} approaches have the best accuracy of our classifiers: an average of $0.5718 \pm 0.0057$ for provinces and an average $0.6224 \pm 0.0054$ for regions. Although the difference with the baseline is not significant for provinces, in both cases we have geographical diversity: the average location accuracy is $0.3896 \pm 0.2268$ for provinces and $0.4765 \pm 0.1794$ for regions.
The \texttt{TF-IDF U} approach under-performs in comparison to the other TF-IDF approaches. This may be explained due to the bias introduced by the high number of user accounts from the most populated places.
The \texttt{LSI U} approach delivers a very low performance, possibly explained by the fact that LSI works best with representative documents and big corpuses.

\begin{table}
\centering
\begin{tabular}{lcc}
\toprule
Approach & Accuracy (prov.) & Accuracy (regions) \\
\midrule
Baseline & $0.5710$ & $0.5841$ \\
TF-IDF U & $0.5028 \pm 0.0029$ & $0.5332 \pm 0.0044$ \\
TF-IDF L & $0.5718 \pm 0.0057$ & $0.6224 \pm 0.0054$ \\
TF-IDF H & $0.5660 \pm 0.0032$ & $0.6075 \pm 0.0034$ \\
LSI U ($k = 200$) & $0.1923 \pm 0.0136$ & $0.2711 \pm 0.0032$ \\
\bottomrule
\end{tabular}
\caption{Evaluation results of our classifiers using a $10$-fold stratified cross validation.}
\label{table:evaluation}
\end{table}

\section{Discussion}
In this section we discuss the results in the context of the research questions presented in Section \ref{sec:intro}.

\emph{Does the participating virtual population (in an event in Twitter) represent the physical population, and to what degree?} 
Our sample holds similar relative sizes of twitter accounts per $1000$ inhabitants for all regions ($2.61 \pm 0.97$), and the logarithms of populations are highly correlated. Given the differences in population seen in Figure \ref{fig:populations}, and the relative rate of accounts per location, we can say that our sample is \textbf{spatially representative} and that it reflects the centralism from the physical population without assuming \emph{demographic representativeness}. In addition, since there are users who were not geolocated (because of invalid, empty or long-tail locations), and because there are people who did not participate in the event, but who are active in Twitter, we consider our sample a \textbf{lower bound }of the entire Chilean population in Twitter.

\emph{Given the content generated by the virtual population, is it possible to identify local content using a lexical approach?} 
According to our results, it is possible to identify local content in spite of the bias towards populated locations using solely content generated by users. This means that \textbf{vocabulary is highly related to geography}. Perhaps it is possible to have better classification results if we incorporate more signals already available into the classifiers, such as user meta-data and available tweet geolocation.

Our results are limited by the following factors:

\textbf{Users do not tweet only about their locations, sometimes not about a location at all}. 
To evaluate the accuracy of our classifiers, we assumed that all tweets by an account are related to its location. This is not accurate. A possible solution is to perform manual labeling and evaluation, but this would be hard, as it requires expertise to be able to identify local content.

\textbf{Dataset is mostly mono-thematic}. 
There are events that generate national-level discussion but are not inherently local, for example, a soccer match of the national team. It is not clear if our approach can be applied successfully in these scenarios. In addition to this, the performance of the LSI classifier draws our attention, in the sense that theoretically LSI is better than plain TF-IDF for language modeling. This result could be an additional signal of the biases present in the dataset.

\section{Conclusions}

In this paper we presented a methodology to normalize social media impact according to geography, i.e., to be able to find and measure \emph{geographically diverse} content. Our methodology is based on the premise that, in addition to a common vocabulary, each location in a country has its own vocabulary subset, comprised of unique hashtags, keywords and names. To test our methodology, we performed a case study in which we analyzed the virtual population of Chile in Twitter that participated in an event of national importance. We found that the virtual population is highly correlated to the physical population and that they share the same spatial distribution. We were able to find the vocabulary that characterized locations, and based on that finding, we built classifiers that predict the location a tweet is referring to with similar performance than a non-geographically diverse baseline at provincial level, and better performance at the regional level. In both cases, our approach has the potential to help to build a geographically diverse timeline and provide diverse recommendations from a geographical perspective.  
Future work should study ways to understand the degree of representation of the virtual population, and generalize the approach to work with multi-thematic streams and other countries.

\section*{Acknowledgments}

{
\small
This work was partially funded by Grant TIN2009-14560-C03-01 of the Ministry of Science and Innovation of Spain. 
}

\small
\bibliographystyle{plain}
\bibliography{chilechi_diversity}

\end{document}